\documentclass{PoS}
\title{A 0535+26 in outburst: magnetospheric instabilities 
and accretion geometry}
\ShortTitle{A0535+26 in outburst}
\author{\speaker{I.~Caballero}$^{a}$, U.~Kraus$^{a}$, 
K.~Postnov$^{b}$, A.~Santangelo$^{a}$, P.~Kretschmar$^{c}$, 
D.~Klochkov$^{a}$ and R.~Staubert$^{a}$ \\
\\{$^{a}$} Institut f\"ur Astronomie und Astrophysik, 
University of T\"ubingen \\ Sand 1, 72076 T\"ubingen, Germany\\
\\{$^{b}$} Sternberg Astronomical Institute, Moscow University\\
119992 Moscow, Russia \\
\\{$^{c}$} ISOC, European Space Astronomy Centre, European Space Agency\\
aaPO Box 78, 28691 Villanueva de la Ca$\tilde{n}$ada, Madrid, Spain\\
E-mail: \email{isabel.caballero@astro.uni-tuebingen.de},\,~\email{kraus@astro.uni-tuebingen.de}, \email{kpostnov@gmail.com},\,~\email{santangelo@astro.uni-tuebingen.de},
\,~\email{Peter.Kretschmar@esa.int},\,~\email{klochkov@astro.uni-tuebingen.de},
\,~\email{staubert@astro.uni-tuebingen.de}}
\abstract{The Be/X-ray binary A 0535+26 showed a normal (type I) outburst
in August/September 2005, which reached a maximum X-ray flux of
400mCrab in the 5-100keV range. The outburst was observed by INTEGRAL
and RXTE. 

The energy of the fundamental cyclotron line has
been measured with INTEGRAL and RXTE at $\sim$45keV. 
Flaring activity was observed 
during the rise to the peak of the outburst. 
RXTE observations during one of these flares
found the energy of the fundamental cyclotron line shifted to a
significantly higher position than during the rest of the outburst,
where it remains constant. 
Also, the energy-dependent pulse profiles during the flare 
differ significantly from the rest of the outburst.
These differences have been interpreted with the presence of
magnetospheric instabilities at the onset of the accretion.

A decomposition method is applied to A 0535+26 pulse profiles.
Basic assumptions of the method are that the asymmetry observed in 
the pulse profiles is caused by non-antipodal magnetic poles, 
and that the emission regions have axisymmetric beam patterns. 
Using pulse profiles obtained from RXTE observations, 
the contribution of the two emission regions has been 
disentangled. Constraints on geometry of the pulsar
and a possible solution of the beam pattern are given. 
First results of the comparison of the reconstructed beam pattern 
with a geometrical model that includes relativistic light deflection 
are presented. }

\FullConference{7th INTEGRAL Workshop\\
		 September 8-11 2008\\
		 Copenhagen, Denmark}
\begin{document}
\section{Introduction}
The Be/X-ray binary A 0535+26 was discovered in 1975 by \cite{rosenberg75}.
It is a transient source, in an eccentric orbit ($e$=0.47) of $\sim110.3$ 
days orbital period \cite{finger96}. Since its discovery it has shown 
5 giant (type II) outbursts in 1980 \cite{nagase82}, June 1982 
\cite{sembay90}, March 1989 \cite{makino89},
February 1994 \cite{finger96} and May 2005 \cite{tueller05}. The last
one in 2005 could not be observed in X-rays by most astronomical
instruments due to Sun constraints. A subsequent normal (type I)
outburst took place in August/September 2005 \cite{finger05} 
associated with the periastron passage. The outburst led to our 
INTEGRAL and RXTE observations. 

The source spectrum presents cyclotron resonance scattering features
at $\sim$45\,keV and $\sim$100\,keV discovered by HEXE
\cite{kendziorra94}. The INTEGRAL and RXTE observations have provided 
the first high accuracy measurements of the cyclotron line energy of A 0535+26. 
RXTE monitoring has allowed to study the evolution of the cyclotron line
energy at different luminosities, and like in other sources 
(\cite{mowlavi06}, \cite{tsygankov06}, \cite{staubert07}) it was found 
to vary with the X-ray luminosity \cite{caballero08_1}.

Flaring activity has been observed during the rise to the peak 
of the outburst. RXTE observations from one of these flares
have revealed a different spectral and timing behavior of the source. 
These results have been presented in \cite{caballero08_1}, and
a possible interpretation of the observed changes is proposed 
in \cite{postnov08}. A review of the observations and interpretation
is given in Sec.~\ref{sec:magnet}.
In Sec.~\ref{sec:profiles} results from a decomposition analysis
of A 0535+26 energy-dependent pulse profiles are shown. 
The method and assumptions are described in Sec.\ref{sec:method}.
A more detailed description is given in \cite{kraus95}, and 
therefore all the formal derivations and technical 
details are not repeated here. 
We find a possible decomposition of the pulse profiles, that has 
allowed to reconstruct the visible part of the beam pattern of A 0535+26, 
and to extract information on the geometry of the system 
(Sec.~\ref{sec:application_A 0535+26}). The reconstructed beam pattern is
used as input for a geometrical model that takes into account 
general relativity effects. Preliminary results are presented in 
Sec.\ref{sec:interpretation}. Sec.~\ref{sec:summary} contains a
summary and conclusions.

\section{Magnetospheric instabilities}
\label{sec:magnet}

\begin{figure}
\includegraphics[height=12cm,angle=90]{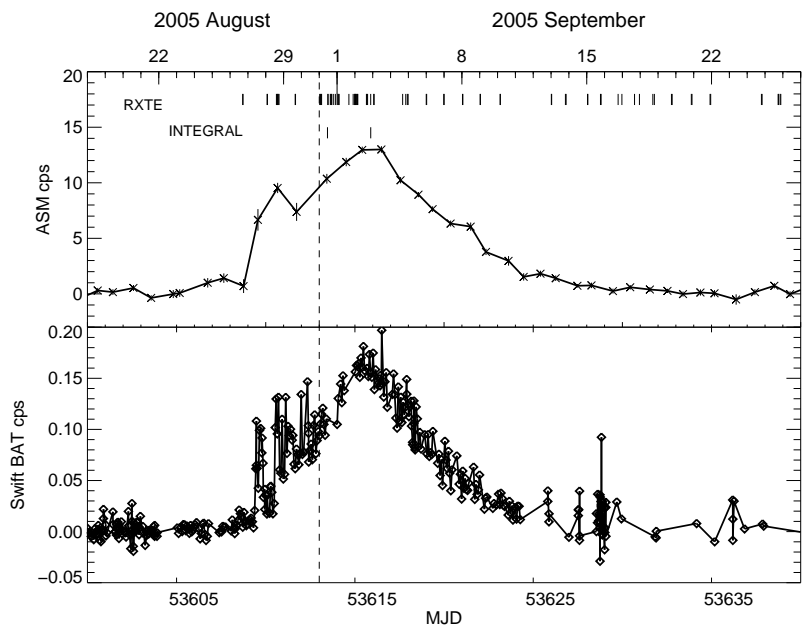}
\caption{A 0535+26 RXTE ASM and Swift BAT light curves during the normal 
(type I) outburst in August/September 2005. The dashed vertical line indicates 
the periastron time. The RXTE and INTEGRAL observations are indicated in 
the upper panel. }
\label{fig:asm_bat}
\end{figure}

Flaring activity has been detected during the rise to the peak of 
the 2005 outburst. RXTE ASM and Swift BAT light curves are shown 
in Fig.~\ref{fig:asm_bat}. INTEGRAL performed a $\sim200\,$ks 
observation near the peak of the outburst, and RXTE 
monitored the complete outburst with an exposure time of 
$\sim140\,$ks, including three observations of one flare 
during the rise to the peak.  The INTEGRAL and RXTE 
observations are indicated in Fig.~\ref{fig:asm_bat}.

The pulse period of the pulsar has been studied in detail during the outburst. 
A constant spin period (within uncertainties) is measured during first 
part of the outburst, $P$=$103.3960(5)\,$s, 
followed by a spin-up of 
$\dot{P}$=$(-1.69\,\pm{0.04})\,\times10^{-8}$ss$^{-1}$  
(measured at MJD 53618) at periastron.
For the first time during a normal outburst a spin-up is measured for 
A 0535+26, providing evidence for the presence of 
an accretion disk \cite{caballero08_1}.

\begin{figure}
\includegraphics[height=6cm]{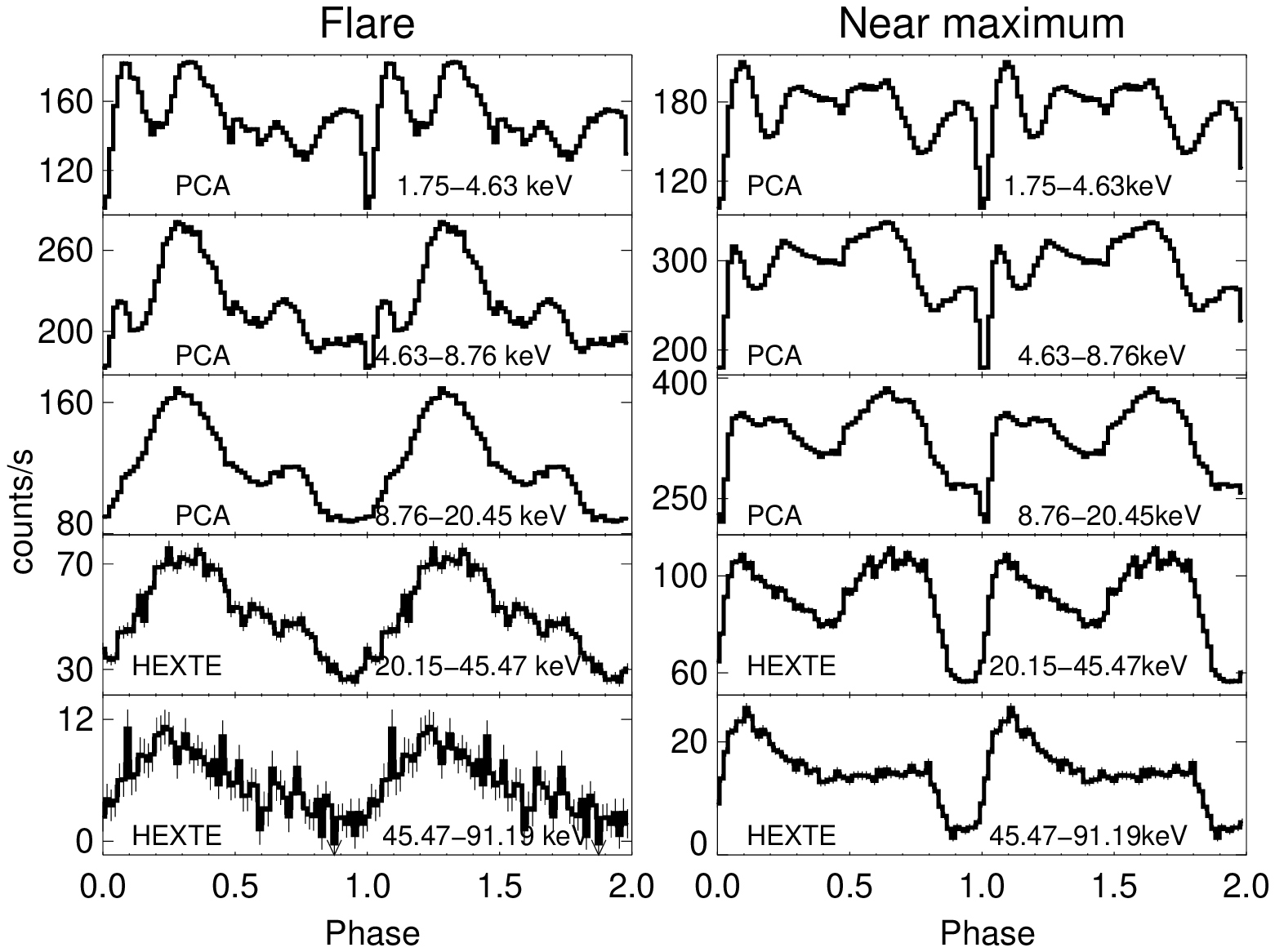}
\includegraphics[height=6cm]{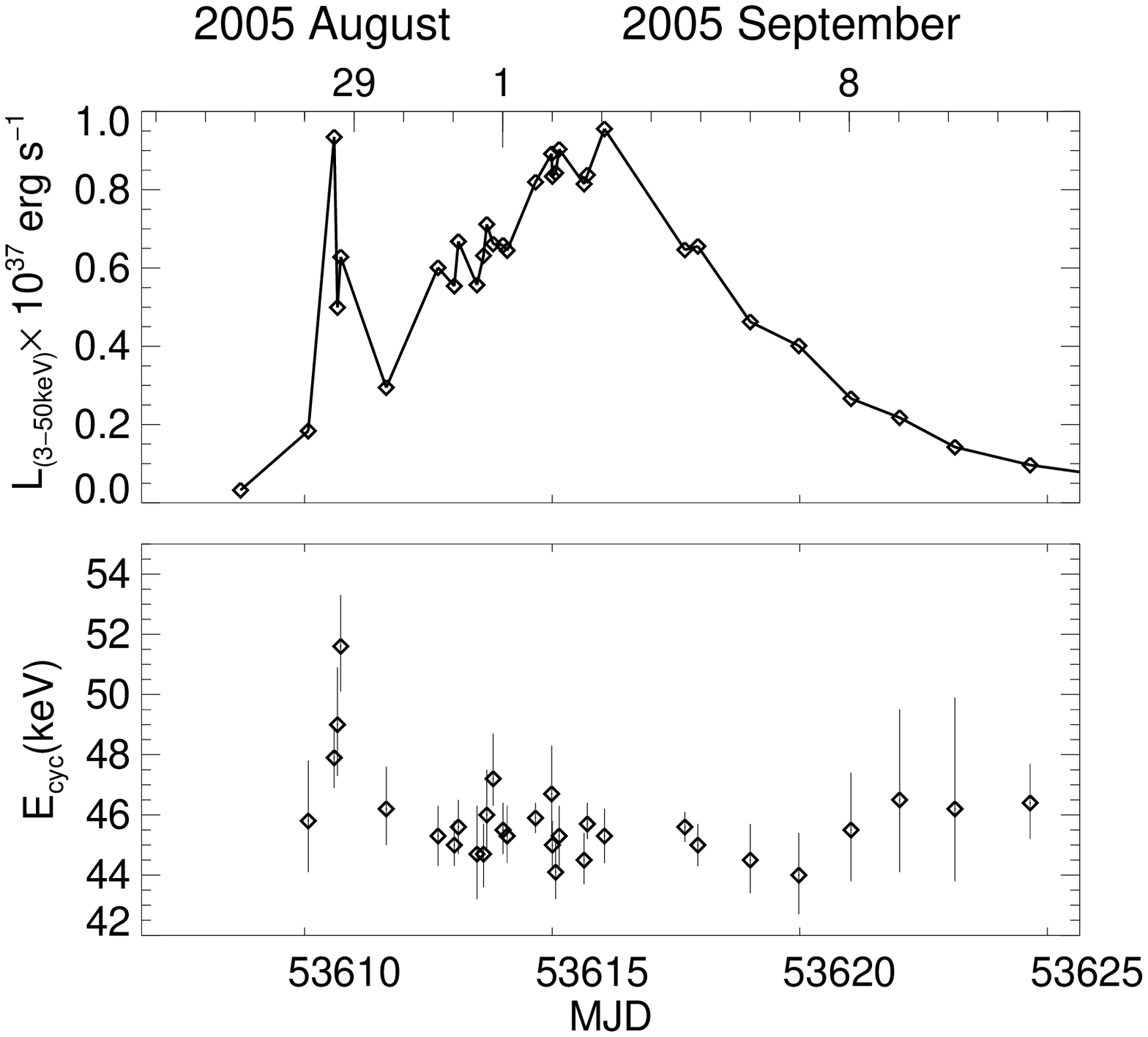}
\caption{Left: A 0535+26 energy-dependent pulse profiles during the flare and near the peak of the August/September 2005 normal (type I) outburst. Right:
 PCA light curve of the outburst (top) and evolution of the cyclotron line
energy (bottom).}
\label{fig:profiles_Ecyc}
\end{figure}

Energy-dependent pulse profiles during the pre-outburst
flare differ considerably from those obtained during
the main part of the outburst, as shown in 
Fig.~\ref{fig:profiles_Ecyc} (left). During the main 
part of the outburst, a strong change of the pulse profiles
above the cyclotron energy takes place. However,
during the flare there is a smooth evolution of the pulse profiles
above the cyclotron energy. The drastic change
above the cyclotron energy during the main peak has been interpreted 
in \cite{postnov08} as a weakening of the fan-like component of the beam 
due to the energy dependence of the cross-section for ordinary photons.
The cyclotron line energy remains constant during the 
main part of the outburst. Near the peak,  INTEGRAL measured
$E_{\mathrm{cyc}}$=$45.9\pm0.3\,$keV \cite{caballero07}. 
A significant change in the cyclotron line is measured 
during the pre-outburst flare, reaching
$E_{\mathrm{cyc}}$=$52.0^{+1.6}_{-1.4}\,$keV. The cyclotron line
evolution during the outburst is shown in  
Fig.~\ref{fig:profiles_Ecyc} (right).

The flaring activity observed has been explained as due to low-mode 
magnetospheric instability that develops at the onset of the accretion, in 
the thin boundary layer between the accretion disk and neutron star 
magnetosphere. The matter accumulated in the 
boundary layer rapidly falls onto the neutron star surface close
to the magnetic poles, along different magnetic field lines by which
stationary accretion proceeds. This explains 
the differences in the pulse profiles and the change in the cyclotron 
energy during the flare compared
to the rest of the outburst. Details are given in \cite{postnov08}.
The present analysis will be performed (and is applicable) only to the 
stable pulse profiles in the main outburst. 

\section{Pulse profile decomposition}
\label{sec:profiles}

Geometrical models of filled and hollow accretion columns of accreting 
neutron stars, including relativistic light deflection, have been 
computed in \cite{kraus01} and \cite{kraus03}. 
These models give the beam pattern or energy-dependent flux of one 
emission region as a function of the angle, 
as seen by a distant observer. Introducing the rotation of the 
pulsar and its geometry, i.e., the orientation of the rotation
axis with respect to the direction of observation and the location of
the two poles, the pulsed emission from each of the two poles  
or single-pole pulse profiles that a distant observer would see 
can be modeled. The sum of the single-pole contributions
gives the total pulse profile.

In this work we apply a decomposition method that inverts the 
process described above. From the observed pulse profile the single-pole
pulse profiles are obtained, and from the single-pole contributions 
information on the geometry of the neutron star and the 
beam pattern is extracted. This method has been applied to the
accreting pulsars Cen X-3 \cite{kraus96} and Her X-1 \cite{blum00}.
Of course the backwards process is not straightforward and involves
ambiguities which will be discussed. 

\subsection{Method}\label{sec:method}

\subsubsection{Assumptions}\label{sec:asumptions}

The basic assumption, which is often adopted in 
model calculations, is that the emission regions at 
the magnetic poles are axisymmetric. 
The beam pattern from one pole is then only a function of the 
angle $\theta$ between the direction of observation and the magnetic axis. 
This makes the single-pole pulse profile necessarily symmetric. One 
of the symmetry points will be the instant when the magnetic axis is closest
to the line of sight, and the other symmetry point will be half a period
later, with the magnetic axis pointing away from the observer. 

Assuming an ideal dipole magnetic field,
the above assumption necessarily implies a symmetric total pulse profile. 
It can be shown that a small displacement of one of the magnetic
poles from the antipodal position can explain
the asymmetry in the sum of the single-pole contributions. 
Therefore the assumption of an ideal dipole field is 
modified, introducing a small offset from an ideal dipole field. 

Another assumption made is that the two emission regions are the same, 
i.e., have the same beam pattern. This implies that each of the two poles
will make visible one section of the same beam pattern. Depending 
on the geometry of the neutron star and the angle of observation, 
those two sections will in some cases have coincident parts.
This assumption has been tested with the accreting pulsars Cen X-3 
\cite{kraus96} and Her X-1 \cite{blum00}. In those cases an 
overlapping region was found, in agreement with the assumption
of two equal emission regions.  

\subsubsection{Decomposition into single-pole pulse profiles}
\label{sec:aliasing}
The first step of the analysis is to express the 
original pulse profile as a Fourier series. The total pulse profile F 
is written as:

\begin{equation}
F(\Phi)=\frac{1}{2}u_{0}+\displaystyle\sum_{k=1}^{n/2-1} [u_{k}\cos(k\Phi)
+v_{k}\sin(k\Phi)]+u_{n/2}\cos(\frac{n}{2}\Phi)
\label{eq:fourier}
\end{equation}

where n is the number of bins of the original pulse profile
and $\Phi$ is the phase. 
Eq.~\ref{eq:fourier} gives a valid representation of the original pulse
profile at all phases if the Fourier transform of F approaches to zero 
as the frequency approaches n/2. If this is not the case (phenomenon called
\emph{aliasing}), less Fourier coefficients are taken into account
to describe the original pulse profile.

The single-pole pulse profiles $f_{1}$ and $f_{2}$ are described
by the following symmetric functions:

\begin{equation}
f_{1}(\Phi)=\frac{1}{2}c_{0}+\displaystyle\sum_{k=1}^{n/2} c_{k}\cos[k(\Phi-\Phi_{1})]
\end{equation}

\begin{equation}
f_{2}(\Phi)=\frac{1}{2}d_{0}+\displaystyle\sum_{k=1}^{n/2} d_{k}\cos\{k[\Phi-(\Phi_{2}+\pi)]\}
\end{equation}

$\Phi_{1}$ and $\Phi_{2}$ are the symmetry points of $f_{1}(\Phi)$ and 
$f_{2}(\Phi)$ respectively. Formally, a decomposition of
F into two symmetric functions exists for every choice of their
symmetry points $\Phi_{1}$ and $\Phi_{2}$. For convenience,
we use the parameter $\Delta := \pi -(\Phi_{1}-\Phi_{2})$,
that represents the azimuthal displacement of one pole with respect 
to the antipodal position (see Sec.~\ref{sec:profiles_to_bp}). 
All formal decompositions will be contained
in the parameter space $\Phi_{1}-\Delta$, with $0\le\Phi_{1}\le\pi$
and $0\le\Delta\le\pi/2$. Once the formal decompositions are calculated, 
physical criteria are applied to decrease the number of decompositions to
physically meaningful ones. Those criteria are the following:

\begin{itemize}
\item ``positive criterion'': both $f_{1}(\Phi)$ and $f_{2}(\Phi)$ 
must be positive, since they represent photon fluxes.
\item ``non-ripples criterion'': the single-pole contributions 
$f_{1}(\Phi)$ and $f_{2}(\Phi)$ should not be much more complicated than 
the original pulse profile. Individual pulse profiles with many peaks 
that cancel out in the sum are not accepted.
\item the same symmetry points must give valid decompositions in all 
energy bands. 
\end{itemize}
Once a possible decomposition is found, the symmetry points
for each of the two poles $\Phi_{1}$ and $\Phi_{2}$, and the 
parameter $\Delta$, related to the position of the emission regions 
on the neutron star, are determined. 
\subsubsection{From single-pole pulse profiles to beam pattern}
\label{sec:profiles_to_bp}

\begin{figure}
\centering
\includegraphics[height=6cm,angle=180,bb=138 720 542 408,clip=]{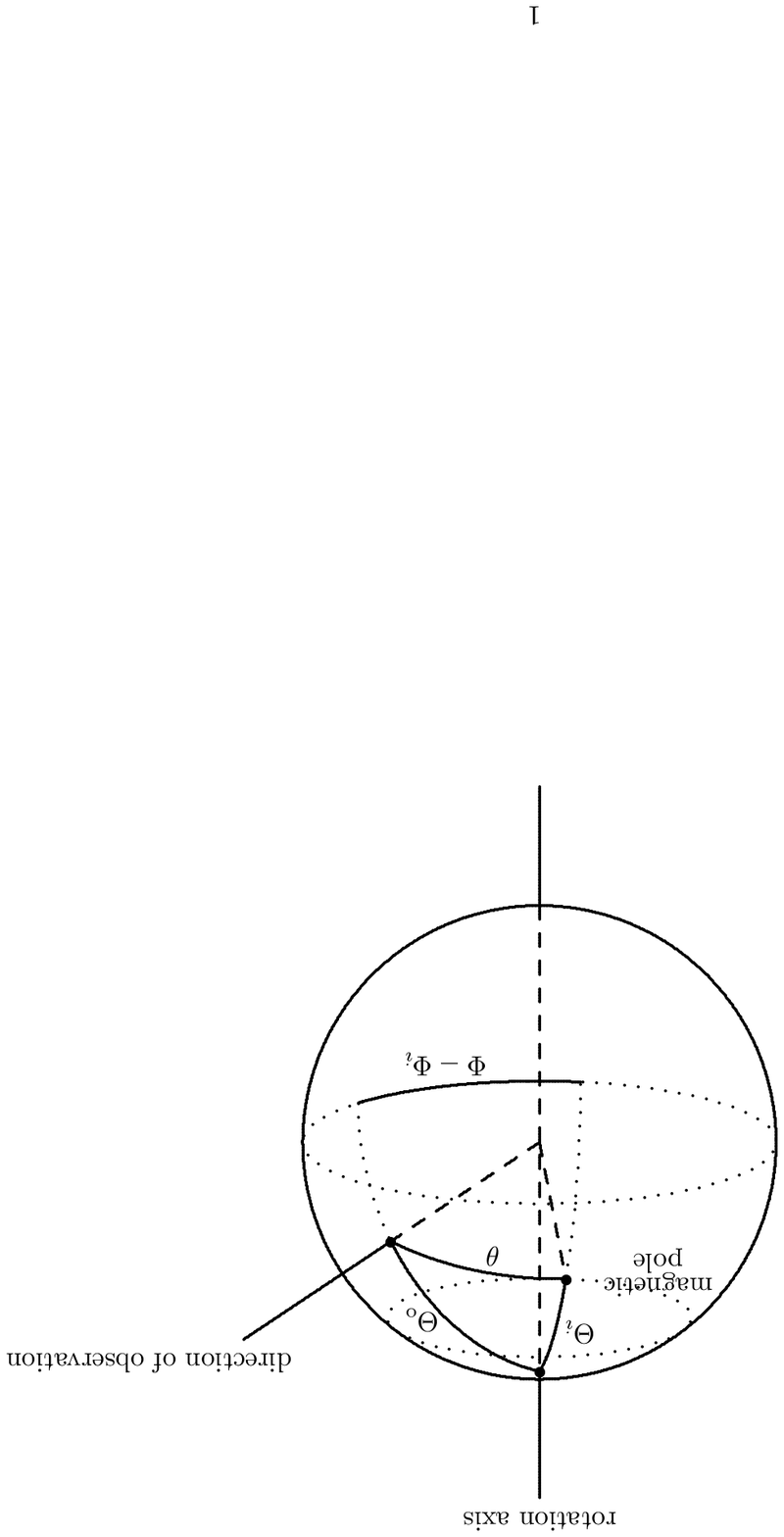}
\caption{Schematic view of a rotating neutron star. A spherical coordinate 
system is used, with the rotation axis as polar axis. $\Theta_{i}$ is the 
polar angle of the \textsl{i}th pole. $\Theta_{0}$
is the polar angle of the direction of observation. The angle $\theta$ 
between the magnetic pole and the direction of observation changes with 
the rotation angle $\Phi$. Figure from \cite{kraus95}.}
\label{fig:kraus95_1}
\end{figure}

In Fig.~\ref{fig:kraus95_1} a schematic view of a rotating neutron 
star is shown. A spherical coordinate system is used with the rotation 
axis as polar axis.  
As was explained above, the beam pattern is assumed to be axisymmetric, 
and therefore to depend only on the angle between the direction of 
observation and the magnetic axis $\theta$. The value of $\theta$ changes 
with the rotation angle $\Phi$. 
Depending on the position of the poles with 
respect to the rotation axis and depending on the direction of observation
with respect to the magnetic axis, we will only observe a section 
of the beam pattern for each pole. 

Applying the cosine formula to the spherical triangle in 
Fig.~\ref{fig:kraus95_1}, we obtain $\theta$ as a function 
of the phase $\Phi$:

     \begin{equation}
     \cos\theta=\cos\Theta_{0}\cos\Theta_{i}+\sin\Theta_{0}\sin\Theta_{i}\cos(\Phi-\Phi_{i})
     \label{eq:theta}
     \end{equation}  

where $\Theta_{0}$ is the polar angle of the direction of observation, 
$\Theta_{i}$ the polar angle of the \textsl{i}th pole, and
$\Phi_{i}$ one symmetry point for the \textsl{i}th pole.

\begin{figure}
\centering
\includegraphics[height=6cm,angle=0]{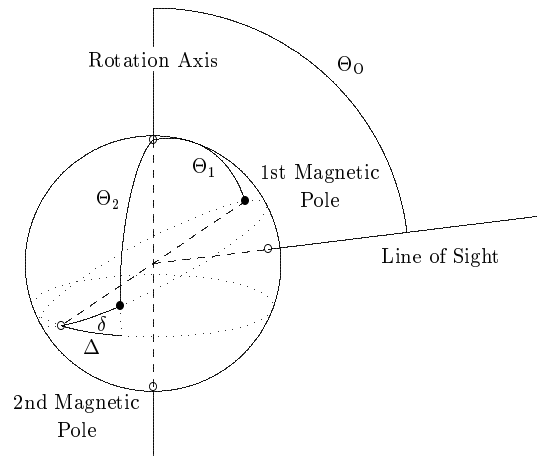}
\caption{Intrinsic geometry of the pulsar. With the rotation
axis as polar axis, the magnetic poles are located at polar 
angles $\Theta_{1}$ and $\Theta_{2}$. The angular distance $\delta$
between the second magnetic pole and the point that is antipodal 
to the first magnetic pole gives the deviation from an ideal dipole
field.
Figure from \cite{blum00}.}
\label{fig:pulsar_geo}
\end{figure}

The intrinsic pulsar geometry is shown in Fig.~\ref{fig:pulsar_geo}.
A complete description of the pulsar can be given in terms
of the polar angles $\Theta_{1}$ and $\Theta_{2}$, and the difference
in their azimuthal angles $\Phi_{1}-\Phi_{2}=\pi-\Delta$. 
The angular distance $\delta$ between the location 
of the second magnetic pole and the point that is antipodal to the 
first magnetic pole can be used as a measure for the deviation 
from an ideal dipole field. From Fig.~\ref{fig:pulsar_geo}:

\begin{equation}
\cos\delta=-\cos\Theta_{2}\cos\Theta_{1}+\sin\Theta_{2}\sin\Theta_{1}\cos\Delta
\label{eq:delta}
\end{equation}

Considering the beam pattern as a function of  $\cos\theta$ and 
the single pole pulse profiles as functions of $\cos(\Phi-\Phi_{i})$,   
as the relation between $\cos\theta$ and $\cos(\Phi-\Phi_{i})$ 
is linear (Eq.\ref{eq:theta}), there is no distortion between the 
two functions. Therefore, once we have the single-pole pulse 
profiles, plotting them as a function 
of $\cos(\Phi-\Phi_{i})$ we will obtain two sections of the beam pattern.
In some cases an overlapping region will emerge. 
Using geometrical properties, it can be shown that 
the relative position of the two single-pole profiles
(in terms of the total pulse profile) are related
to the position of the two poles.
This requires an independent determination of the direction 
of observation. Details can be found in \cite{kraus95}.

\subsection{Application to A 0535+26}
\label{sec:application_A 0535+26}
RXTE HEXTE energy-dependent pulse profiles of A 0535+26 
during the August/September 2005  
normal outburst have been analyzed.
Pulse profiles obtained during the main part of the outburst
have been selected for the analysis, as those profiles
appear to be very stable, not only during the outburst
but also compared to historical observations. 
Several observations have been used, for a total 
exposure time of $\sim$20\,ks.

\subsubsection{Search for acceptable decompositions}
\begin{figure}
\includegraphics[height=5.5cm,angle=0]{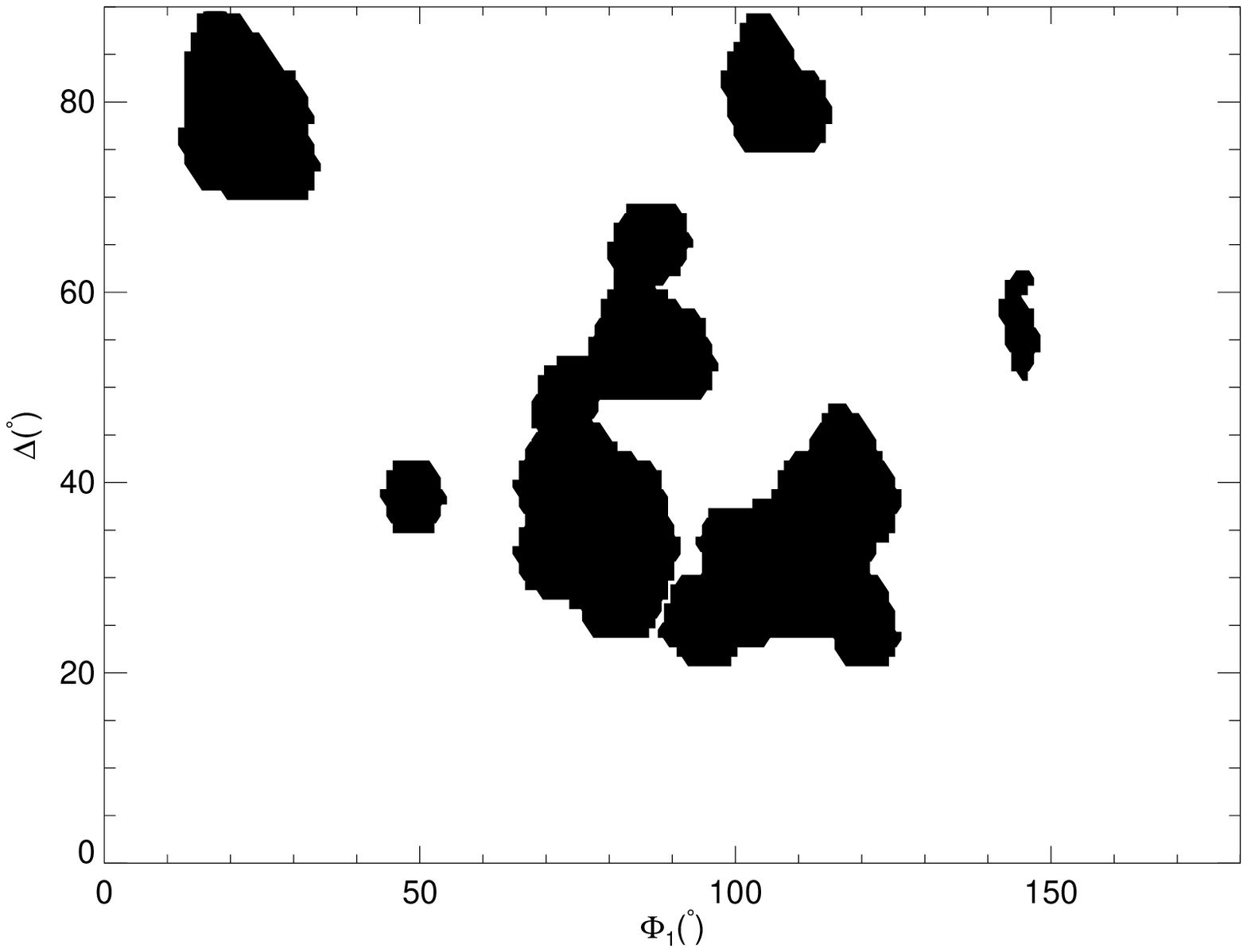}
\includegraphics[height=5.5cm,angle=0]{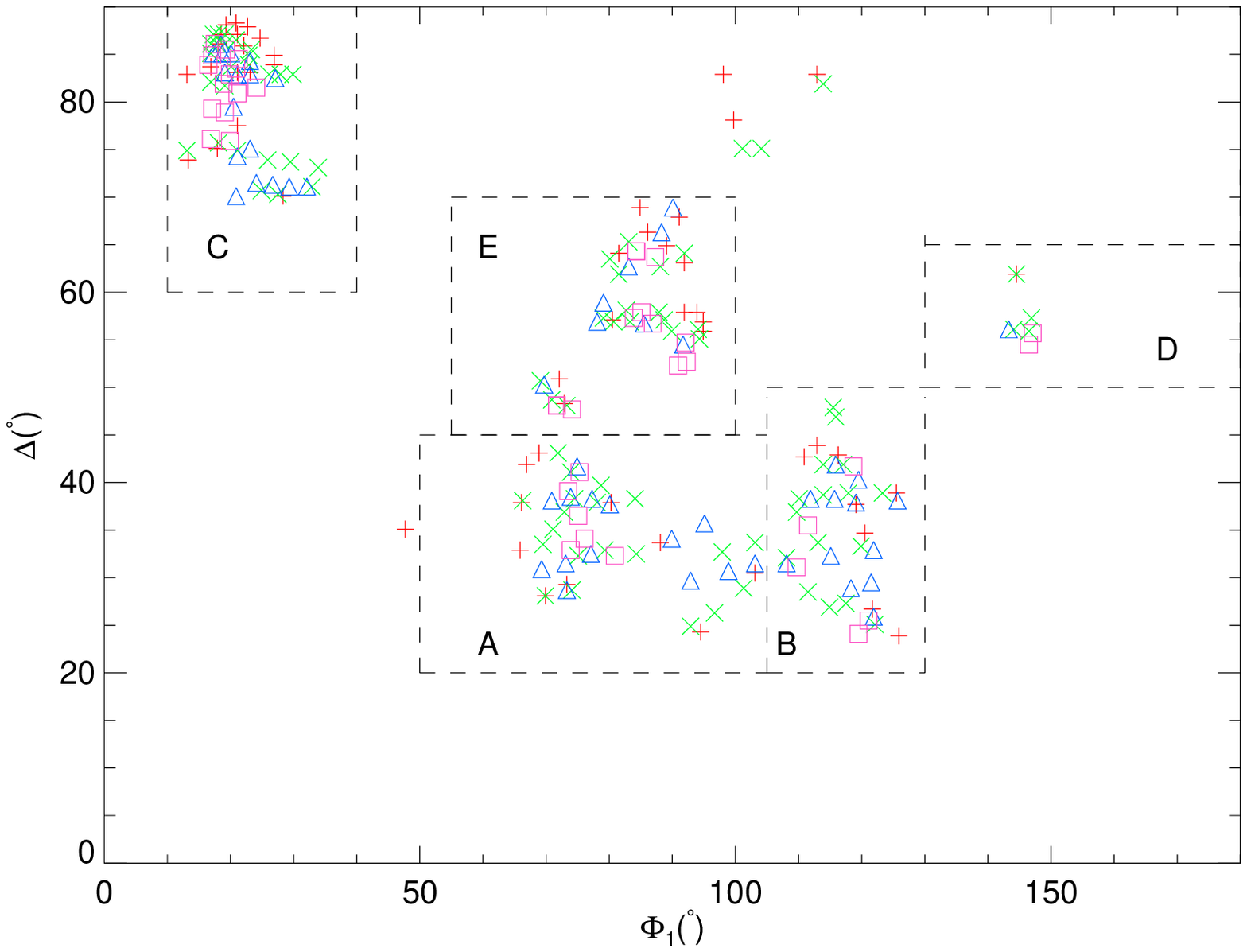}
\caption{Left: Result of applying the ``non-negative'' criterion: 
decompositions outside the black area are discarded. Right: 
Highest ranked decompositions grouped in five regions 
$A$, $B$, $C$, $D$ and $E$. 
The different symbols represent different energy ranges: 
+: (18.3--30.9)\,keV,
$\times$: (30.9--44.5)\,keV , 
$\triangle$: (44.5--59.1)\,keV, 
$\Box$: (59.1--99.8)\,keV }
\label{fig:contour_decompositions}
\end{figure}

The energy-dependent pulse profiles are written as Fourier series.
These functions are then written as the sum of two symmetric functions 
$f_{1}(\Phi)$ and $f_{2}(\Phi)$. 
To search for physically meaningful decompositions, the $\Phi_{1}-\Delta$ 
parameter space is divided in $1\,^{\circ}\times1\,^{\circ}$ boxes. All the
formal decompositions are represented in this plane. 
We apply the ``non-negative criterion''. Out of all the decompositions, only
those for which $f_{1}(\Phi)$ and $f_{2}(\Phi)$ are positive are accepted. 
Fig.~\ref{fig:contour_decompositions} (left) shows where in the 
$\Phi_{1}-\Delta$ plane positive decompositions have been found. 
The ``non-ripples criterion'' is then applied. A quality 
function is defined, which counts the number of 
peaks in one profile. To avoid having to study all the decompositions, similar
ones are grouped into types. We study the representative of each type.
Fig.~\ref{fig:contour_decompositions} (right) 
shows where the highest ranked profile representatives were found. 
We have examined all these decompositions in different energy ranges,
dividing the parameter space in five regions $A$, $B$, $C$, $D$ and $E$.

\begin{figure}
\includegraphics[height=8cm,angle=0]{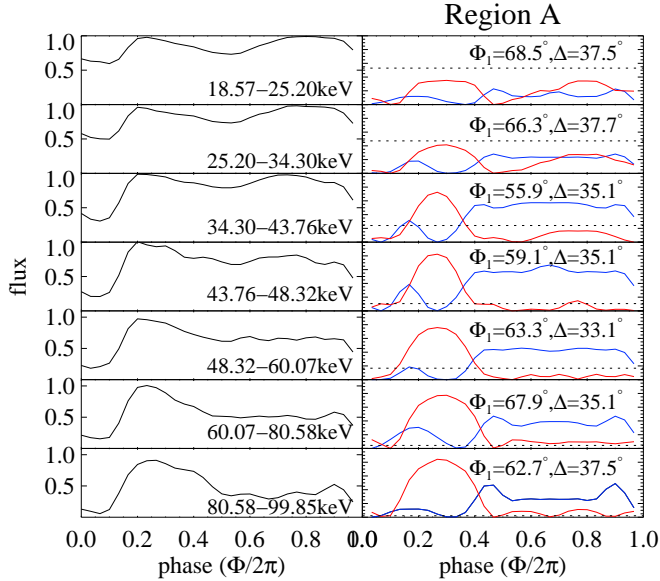}
\caption{Best decomposition of the original pulse profiles (left) in two symmetric functions (right).}
\label{fig:decomp}
\end{figure}

Combining the results of the analysis in different energy ranges and
for different observations, we find a best decomposition of the 
original pulse profiles in region A. 
They are shown in Fig.~\ref{fig:decomp}.
The dotted horizontal line corresponds to the remaining unmodulated 
flux available to distribute between the 
two symmetric functions that can not be determined from the 
decomposition. The minima of the two symmetric functions have been shifted
to zero, so that the sum of the two symmetric functions plus
the unmodulated flux reproduces the original pulse profile. 
The values of $\Phi_{1}$, $\Delta$ and $\Phi_{2}$ (using
$\Delta=\pi-(\Phi_{1}-\Phi_{2})$) can not be more accurate
than the $12^{\circ}$ bin resolution, and we therefore use this value
as uncertainty for the symmetry points. 
Average values for the best decomposition in region $A$ are
$\Phi_{1}=73^{\circ}\pm12^{\circ}$, $\Delta=33^{\circ}\pm12^{\circ}$
and $\Phi_{2}+\pi=106^{\circ}\pm12^{\circ}$ 

Decompositions in regions $C$ and $E$ are discarded because
they present a very strong anti-correlation in the main peaks 
which seems artificial, not expected from two independent emission regions. 
Decompositions in regions $B$ and $D$ have also been discarded, 
because the single-pole pulse profiles present an anti-correlation
in many small features that cancel out in the sum, also 
not expected from two independent emission regions. 
Another argument to reject decompositions in regions $C$, $D$ and $E$ is that
they all present higher values of $\Delta$. Under the assumption of 
slightly displaced magnetic poles, smaller values of $\Delta$ are more likely
to be real. This was the case in the analysis of the accreting pulsars 
Cen X-3 \cite{kraus96} and Her X-1 \cite{blum00}, where the 
best decompositions were found for small values of $\Delta$.
A further argument against decompositions in regions
 $B$, $C$ and $D$ emerges in the reconstruction of the beam pattern 
from the single-pole contributions 
(see Sec.~\ref{sec:geo_beam_pattern_A 0535+26}).

\subsubsection{From single-pole pulse profiles to geometry and beam pattern}
\label{sec:geo_beam_pattern_A 0535+26}

\begin{figure}
\includegraphics[height=7cm,angle=0]{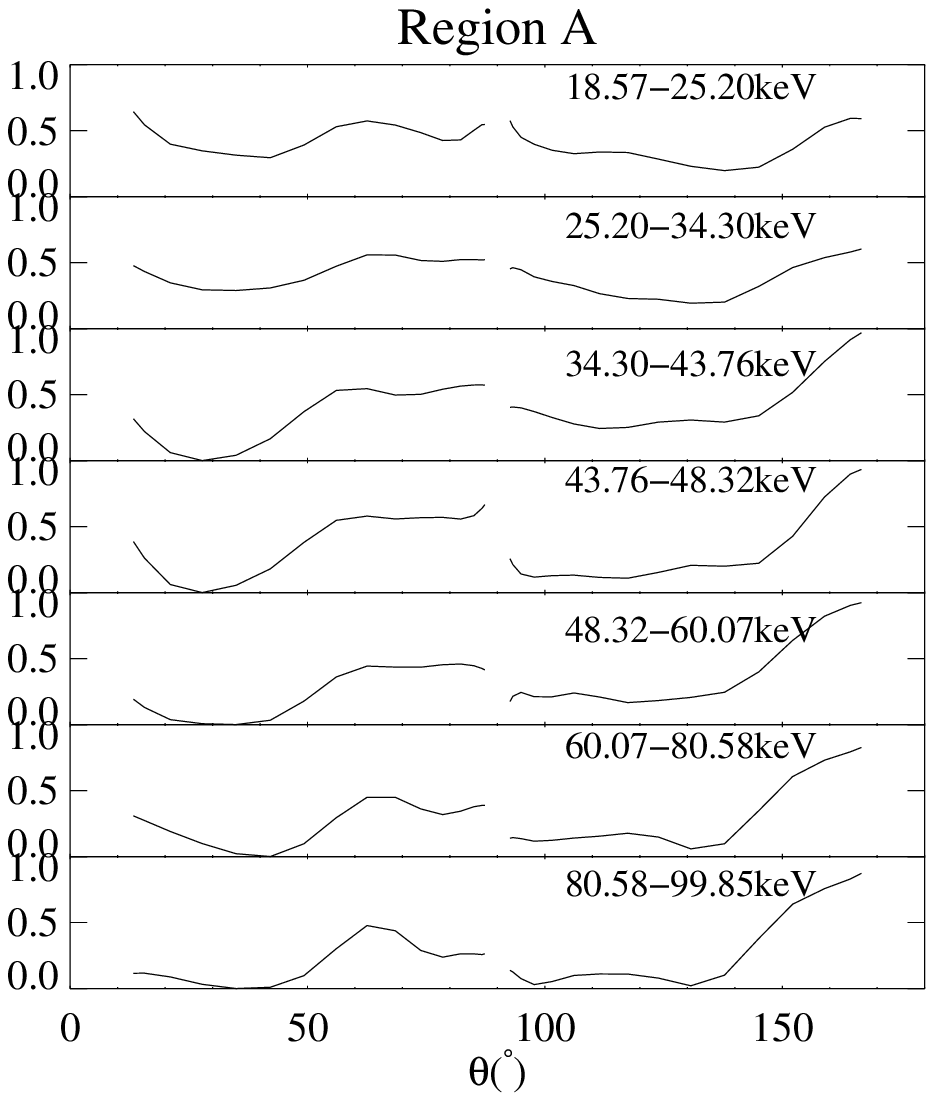}
\includegraphics[height=7cm,angle=0]{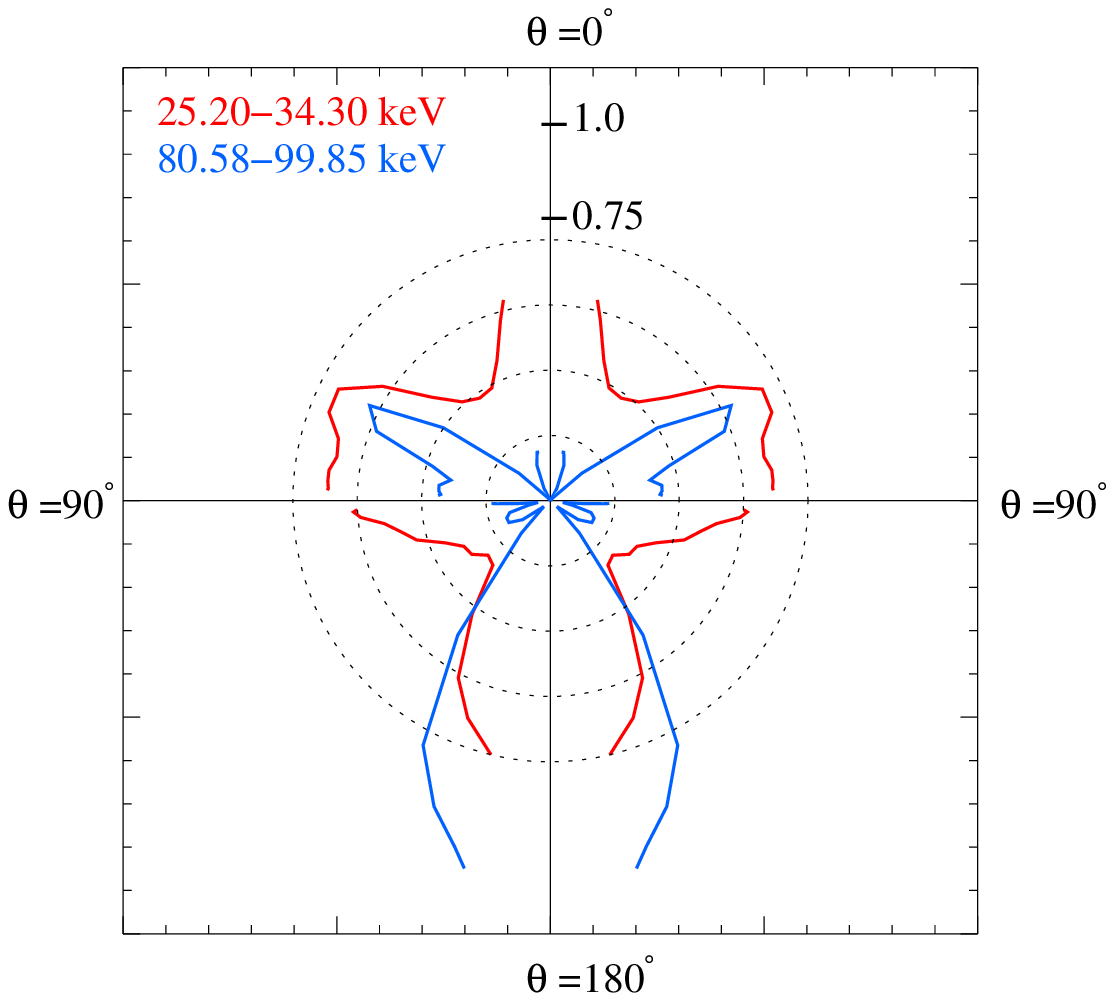}
\caption{Reconstructed beam pattern of A 0535+26 in different energy ranges. 
Left: linear representation. Right: polar diagram.}
\label{fig:bp}
\end{figure}

Doing the proper transformation of the single pole pulse profiles into
undistorted sections of the beam pattern, an overlapping region 
does not emerge. However, under the assumption of 
two identical emission regions, the two sections 
can almost be connected to each other, with a small gap 
in between. This was not possible for the decompositions in 
regions $B$, $C$ and $D$.
The shift between the two sections is related to the position
of the magnetic poles and the direction of observation (see
\cite{kraus95} for details).
The inclination of the system is $i=37\,\pm2\,^{\circ}$ 
\cite{giovannelli07}. Assuming that the rotation axis of the neutron 
star is perpendicular to the orbital plane, 
 $i=\Theta_{0}$. We can therefore obtain the location 
of the poles $\Theta_{1}$ and $\Theta_{2}$. 
The angular distance between the location of the second
pole and the point that is antipodal to the first pole $\delta$
can be estimated using Eq.~\ref{eq:delta}.
The estimated values for the position of the magnetic
poles and the offset are
$\Theta_{1}\approx50^{\circ}$, $\Theta_{2}\approx130^{\circ}$,
$\delta\approx25^{\circ}$.
It is then possible to plot the reconstructed sections of the beam 
pattern as a function of $\theta$. The two sections of the beam 
pattern are reconstructed for $\theta\in(13^{\circ}-87^{\circ})$ and 
$\theta\in(93^{\circ}-167^{\circ})$. The reconstructed beam 
pattern is shown in Fig.~\ref{fig:bp}.

\subsubsection{Interpretation of beam pattern}
\label{sec:interpretation}

A characteristic feature of the reconstructed beam pattern is
a minimum observed in the flux between $\theta\approx30^{\circ}-40^{\circ}$.
This feature is present at all energies and we therefore believe that
it could be related to the geometry of the accretion. (It is also
present at lower energies, seen from the analysis of PCA data 
\cite{caballero08_2}). Filled column models
give a beam pattern in which the flux decreases 
at low values of $\theta$, corresponding to the instant when the observer 
looks along the accretion stream \cite{kraus03}. Introducing a 
hollow column plus a halo created on the neutron star surface 
around the column walls from 
scattered radiation emitted from the walls would explain the flux at 
low values of $\theta$ and the minimum as $\theta$ increases when the
observer looks directly into the column. 

The steep increase in flux at high values of theta ($\theta>120^{\circ}$)
could be due to gravitational light bending, which produces a 
similar feature in model calculations. 


To obtain estimates on the size of the accretion column, a model
for a hollow column plus a halo has been calculated for 
$\theta\in(0^{\circ},40^{\circ})$.
Geometrical models of filled columns, including
the formation of a halo around the accretion column, 
were presented in \cite{kraus03}, where the 
relative importance of the different components 
(halo-column) to the observed flux was studied. 
In the work presented here, the modelling is performed 
as in \cite{kraus03}, but introducing a hollow column. 
A detailed study of this model will be presented elsewhere. 
Beam patterns are computed using ray-tracing \cite{foley90} and
include relativistic light deflection \cite{nollert89}.

\begin{figure}
\includegraphics[height=6cm,angle=0]{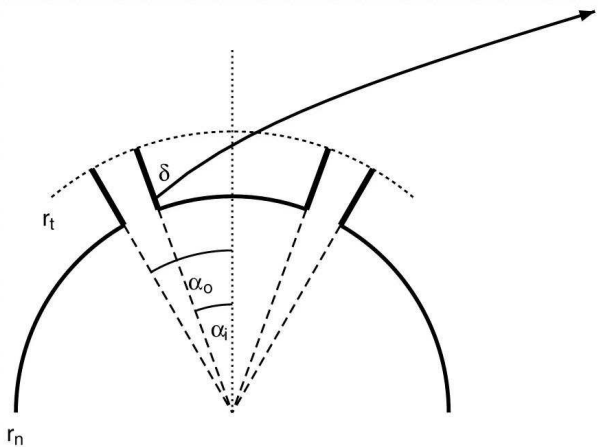}
\caption{Geometrical model of hollow column. Figure from \cite{kraus01}}
\label{fig:geo_model}
\end{figure}

The emission of the column wall is assumed to be a blackbody of
temperature $T_{eff}$, assumed to be isotropic. 
Part of the radiation that leaves the column hits the neutron star 
surface, creating a luminous halo around the column walls. This 
emission is assumed to be thermal. 
Schwarzschild metric is used to calculate photon paths. This
is appropriate for A 0535+26 because it is a slowly rotating neutron star
($P_{spin}\sim103.4\,s$). The Schwarzschild radius is $r_{S}=2GM/c^{2}$
and the neutron star radius $r_{n}$.
The accretion funnel has an
inner half-opening angle $\alpha_{i}$ and outer half-opening 
angle $\alpha_{0}$. A radiative shock is assumed to  
form close to the neutron star surface at radial coordinate $r_{t}$. 
See Fig.~\ref{fig:geo_model} for a sketch of the geometry of the 
hollow column. Below
the shock, the column is optically thick and radiation is emitted
from the inner and outer walls. Above the shock, the accretion is 
assumed to be in free-fall.

The mass and radius of the neutron star are taken to be
$M_{n}=1.4\,$M$_{\odot}$ and $r_{n}=10\,$km. The radiative
shock is formed at $r_{t}=10.5\,$km, and an
asymptotic luminosity per pole of $L_{\infty}=0.8\times10^{37}$ergs$^{-1}$
is taken. The cyclotron energy, used to calculate
the magnetic scattering cross-section,  is set to $E_{\mathrm{cyc}}$=45\,keV
at the magnetic pole. Beam patterns have been computed for 
7.6\,keV photons. 

\begin{figure}
\includegraphics[height=5cm,angle=0]{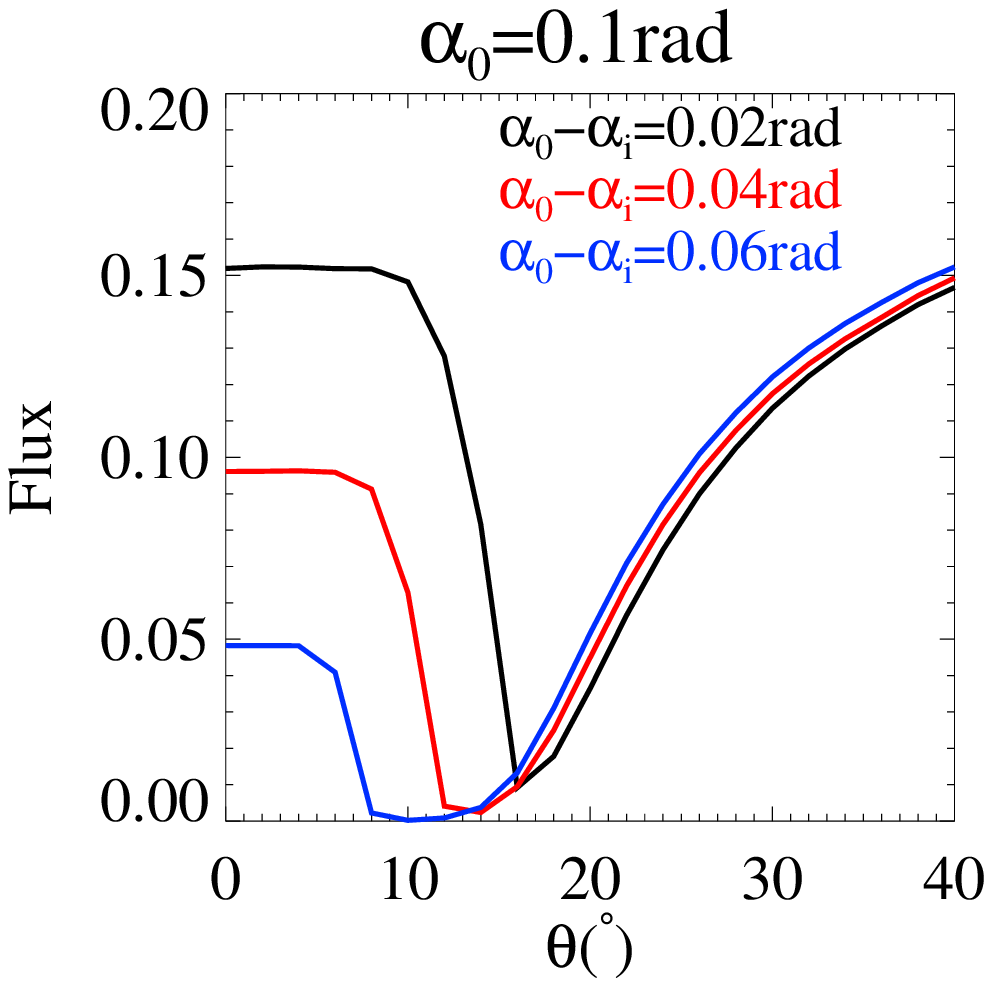}
\includegraphics[height=5cm,angle=0]{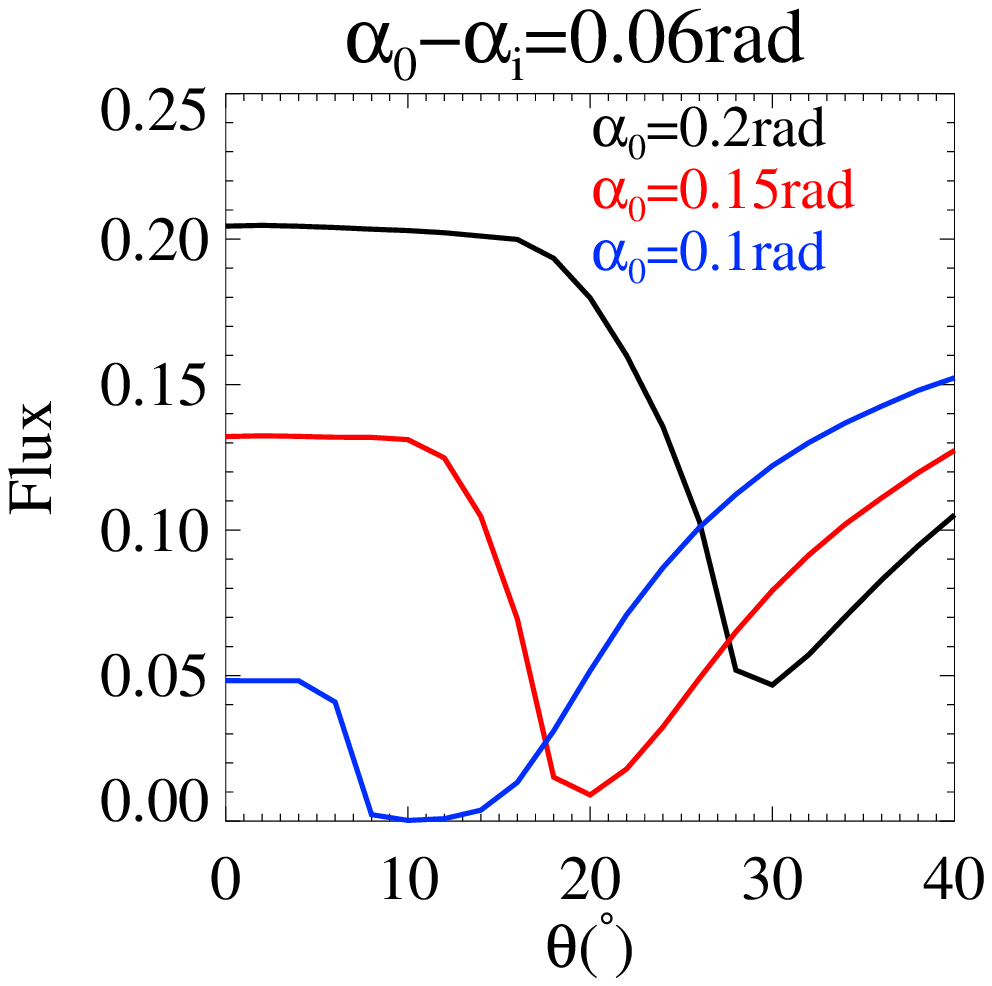}
\caption{Beam pattern models for hollow column with halo 
for different column thickness and the same opening angle (left) 
and different opening angles with same column thickness (right), 
calculated for $\theta\in(0^{\circ},40^{\circ})$.}
\label{fig:diff_models}
\end{figure}

Different models have been computed for different values
of the inner and outer half-opening angles.
For each model, the temperature of the Plank spectrum 
emitted by the column wall and the
density that the accreting material has at the base 
of the free-fall section follow from the assumed 
geometry and the value of $L_{\infty}$. 
The parameters used for each model are listed in Table~\ref{tab:models}.
Results from the model calculations are shown in 
Fig.~\ref{fig:diff_models}.

\begin{table}[h]
\caption{Model parameters used in the computation of beam patterns
for A~0535+26, for $\theta\in(0^{\circ},40^{\circ})$.}
\label{tab:models} \centering
\begin{tabular}{ccccc}\hline
model & $\alpha_{i}$ (rad)& $\alpha_{0}$ (rad) & $kT$(keV)  & $\rho$ ($10^{-5}$g/cm$^{-3}$)\\\hline
1    &      0.08    &  0.1  &4.1  & 16  \\
2    &      0.06    &  0.1  &4.1  & 9 \\
3    &      0.04    &  0.1  &4.1  & 6.8 \\
4    &      0.09    &  0.15 &3.7  & 4 \\
5    &      0.14    &  0.2  &3.5  & 2.8 \\\hline
\end{tabular}
\end{table}

By comparing the A~0535+26 reconstructed beam pattern 
(Fig.~\ref{fig:bp}) with the models in Fig.~\ref{fig:diff_models}, 
it can be seen that the shape of the models reproduce
well the shape of the reconstructed beam pattern. 
We can estimate the half-opening angle and column thickness 
to be $\alpha_{0}=0.2$rad$\sim11.5^{\circ}$ and 
$\alpha_{0}-\alpha_{i}=0.06$rad$\sim3.4^{\circ}$.
Those are preliminary results from an ongoing work that will be presented in 
\cite{caballero08_2}. 

\section{Summary and conclusions}
\label{sec:summary}
In this work results 
from spectral and timing analysis of A 0535+26 during a flare that took
place in the rise to the peak of the August/September 2005 normal outburst
compared to the rest of the outburst have been summarized. 
The differences observed in the pulse profiles and in the cyclotron 
energy can be explained with magnetospheric instabilities that develop
in the onset of the accretion. Those results and interpretation
can be found in \cite{caballero08_1} and \cite{postnov08} respectively.

In the second part of the work, a decomposition analysis has been 
applied to the A 0535+26 energy-dependent pulse profiles. 
A dipole magnetic field is assumed with axisymmetric emission regions. 
The asymmetry in the total pulse profiles is explained with a small 
offset from one of the magnetic poles from the antipodal position. 
We find a physically acceptable decomposition of the pulse profiles 
that allows to extract information on the geometry of the pulsar. 
We obtain $\Theta_{1}\approx50^{\circ}$ and $\Theta_{2}\approx130^{\circ}$ 
for the position of the magnetic poles, and an offset of 
$\delta\approx25^{\circ}$.
The visible section of the beam pattern has been reconstructed. 
First results from a geometrical model of hollow column that includes the
formation of a halo are used to interpret the main 
features of the reconstructed beam pattern of A~0535+26. 
Further details will be presented in \cite{caballero08_2}.



\end{document}